\documentclass[aps,prl,twocolumn,showpacs,groupedaddress]{revtex4} 
\usepackage{graphicx} 
 
\begin{document} 
 
\title{Optical gain removes absorption and increases resolution in a near-field lens} 
\author{S. Anantha Ramakrishna} \author{J.B. Pendry} 
\affiliation{The Blackett Laboratory, Imperial College, London SW7 2BZ, U.K.} 
\date{\today} 
 
\begin{abstract} 
A recent paper showed how to construct a lens that focuses near field 
radiation and hence produce resolution unlimited by wavelength. The 
prescription requires lossless materials with negative refractive index: 
finite loss cuts off the finer details of the image. In this paper we 
suggest compensating for losses by introducing optical gain media into the 
lens design. Calculations demonstrate a dramatic improvement in performance 
for a silver/gain composite medium at optical frequencies. 
 
\end{abstract} 
 
\maketitle


 
There are two sorts of electromagnetic radiation: near field and far field. 
The latter propagate as plane waves with a real wave vector, the former has 
an imaginary wave vector resulting in exponential decay and therefore is 
confined to the vicinity of the source. Conventional lenses act only on the 
far field: focussing the near field requires amplification.  Unfortunately 
for imaging purposes the finer details of an object are contained in the 
near field. Recently it was shown \cite{pendry00} how a `perfect lens' could 
be designed in which both the near and far fields could be persuaded to 
contribute to an image. This perfect lens essentially amplifies 
the near field through a series of surface 
plasmon resonances. In a lossless system this process of amplification 
requires no energy input, other than that from the source, but in the 
presence of losses the performance of the lens rapidly degrades as the 
quality factor of the resonances deteriorates. Losses are the ultimate 
limiting factor for resolution and even a highly conducting metal such as 
silver has  restricted performance as a near-field lens material. In this letter we 
explore the possibility of compensating for the losses in silver by 
constructing a composite material consisting of alternate slices of silver and 
an optical gain material. Our conclusions are that considerable improvement 
in performance is possible by this means. 
 
The electromagnetic field in the 2-D object (x-y) plane can be conveniently 
decomposed into the Fourier components $k_x$ and $k_y$ and polarization 
defined by $\sigma $: 
\begin{eqnarray} 
E(x,y,z;t) &=&\sum_{k_x,k_y,\sigma }E(k_x,k_y,k_z)\times   \nonumber \\ 
&\times &\exp [i(k_xx+k_yy+k_zz-\omega t)], 
\end{eqnarray} 
where the source is assumed to be monochromatic at frequency $\omega $, $%
k_x^2+k_y^2+k_z^2=\omega ^2/c^2$ and $c$ is the speed of light in free 
space. Obviously when we move out of the object plane, the phase of the 
propagating radiative components (real $k_z$ for $k_x^2+k_y^2<\omega ^2/c^2$%
) and the amplitude of the non-propagating evanescent components (imaginary $%
k_z$ for $k_x^2+k_y^2>\omega ^2/c^2$) change, and the image gets blurred. 
The \textit{`perfect lens'} suggested by \cite{pendry00} is just a slab of 
negative refractive material(NRM), where both the dielectric permittivity ($%
\varepsilon $) and the magnetic permeability ($\mu $) are negative 
simultaneously ($n=-\sqrt{\varepsilon \mu }$ in this case). In 1968, 
Veselago \cite{veselago} had pointed out that such a slab would act 
as a lens in that it would refocus the rays from a point source on 
one side to a point image on the other side due to a modified Snell's 
law for negative refractive media. But the fact that the slab also 
acts on the evanescent near-field modes was not realised until 
recently. The perfect lens performs the dual function of correcting 
the phase of the radiative components as well as amplifying the 
near-field components bringing them both together to make a 
perfect image and thereby eliminating the diffraction limit on 
the image resolution.  
 
In general the conditions under which this perfect imaging 
occurs are :  
\begin{equation} 
\varepsilon _{-}=-\varepsilon _{+},~~~~~~~~~~\mu _{-}=-\mu _{+}, 
\end{equation} 
where $\varepsilon _{-}$ and $\mu _{-}$ are the dielectric 
permittivity and magnetic permeability of the NRM slab, and 
$\varepsilon _{+}$ and $\mu _{+}$ are the dielectric permittivity 
and magnetic permeability of the surrounding medium respectively.  
Under these conditions the transmission coefficient of the slab 
is exactly $\exp (-ik_zd)$, where $d$ is the thickness of the slab. 
We have earlier pointed out that these are precisely the conditions 
for the existence of surface plasmon modes 
at the surface \cite{pendry00,anantha02a} and that it is 
sufficient to meet these conditions at any one interface to enable 
the amplification of evanescence in the slab \cite{anantha02a}. 
Although Veselago's early study of negative refraction was entirely 
speculative, the topic has become important as the media with 
negative refraction at microwave frequencies can now be physically 
realised \cite{pendry96,pendryIEEE,smith00,smith01}. The 
physics of negative refractive index has now caught the 
imagination of the physics community as evidenced by the 
recent publications \cite 
{smith00,smith01,ruppin,soukoulis01,tretyakov01,itoh01,solymar,zhang02,steve02,smithprep}%
. 
  
As it is difficult to obtain negative permeable media at optical 
frequencies, it was suggested in the original letter \cite{pendry00} 
that metals (such as silver) with negative permittivity alone 
could be good lens materials for p-polarized light. Particularly 
in the near field limit, where $k_x^2+k_y^2 \gg \omega^2/c^2$, 
the electric and magnetic fields are independent of one another, 
if we confine our attentions to electric fields we have a 
simplified requirement for the function of the lens, 
\begin{equation} 
\label{PLeps_eqn} 
\varepsilon _{-}=-\varepsilon _{+}, 
\end{equation} 
with the magnetic permeability now becoming irrelevant.  The 
performance of a silver lens is, however, 
limited by losses and the positive magnetic permeability. Their 
effects can be reduced by redesigning the lens as a multilayer 
stack of very thin alternating layers of metal (or NRM) and positive 
dielectric medium, and obtain good sub-wavelength image resolution 
\cite{anantha02b,solymar}. This system has the disadvantage that the 
object and image planes are very close to the edges of the stack, 
but can be used to transfer the image from one point to another 
in the manner of a `fibre-optic bundle'. However even in this 
system losses eventually limit the resolution.  In equation 
(\ref{PLeps_eqn}) above we see the possibility that if 
$\varepsilon _{-}$ is lossy, i.e.  
\begin{equation}
\mathrm{Im} (\varepsilon _{-})>0, 
\end{equation}
then the condition may still be satisfied provided that  
\begin{equation}
\mathrm{Im}(\varepsilon _{+})<0. 
\end{equation}
In other words we need a medium which exhibits gain. We explore the 
possibility of using optical amplification/gain in the positive 
dielectric medium to coherently compensate for the absorption in 
the negative dielectric medium. This enables the design of a 
`fibre-optic bundle' which consists of a multilayer stack of 
alternating thin layers of lossy metal (silver) and an amplifying 
positive dielectric medium (an optically pumped semiconductor, 
for example) to transfer images with good sub-wavelength resolution 
across large stack thicknesses (of the order of a few wavelengths). 
Such a fibre-optic probe has the novel property of acting on the 
near-field evanescent components as well as the radiative 
components of a source. 
 
We approach this by considering the perfect-lens condition on the 
dielectric permittivity in Eq. (\ref{PLeps_eqn}). Now the 
dielectric permittivity of a causal negative dielectric medium 
would both disperse with frequency and also be dissipative. 
The dissipation which shows up as an imaginary part of the 
dielectric constant represents a deviation from the perfect lens 
condition, and limits the image resolution. To satisfy 
the perfect lens conditions in 
both the real and the imaginary parts, we would require  
\begin{equation} 
\varepsilon _{+} = \varepsilon _{+}^{^{\prime }}- i\varepsilon_{+}^{^{\prime 
\prime }},~~~~~~~~~~~ \varepsilon _{-} = -\varepsilon_{+}^{^{\prime }}+ 
i \varepsilon _{+}^{^{\prime \prime }}. 
\end{equation} 
In other words, the positive medium should be optically amplifying (as in a 
laser gain medium) in order to counter the effects of absorption in the 
negative medium. Then the perfect lens conditions would be `perfectly' met 
and the perfect image would result, at least in the quasi-static limit. Now 
the surface states at the interface between the dissipative metal and an 
amplifying positive dielectric displays a very interesting behaviour. There 
is a net flux of energy across the interface - from the positive amplifying 
medium into the absorbing negative medium. This is in contrast to the 
lossless (and gainless) system, where there is no net flux of energy normal 
to the interface and the fields are purely evanescent.
We should also note that such a cancellation 
of absorption in one location by generation in another is possible only 
because both absorption and amplification do not cause dephasing of the wave  
--a consequence due to the bosonic nature of light that permits  
stimulated absorption and stimulated emission. 
\begin{figure}[tbp] 
\includegraphics[width=8cm]{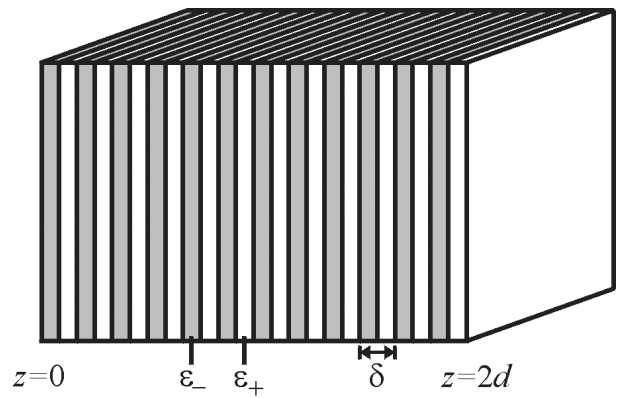} 
\caption{Schematic of the layered structure considered here. The positive 
amplifying and negative dissipative dielectric layers are assumed to be of 
equal thickness $\delta/2$. The object and image planes are on either side 
of the stack at a distance of $\delta/4$ from the edges. The total length of 
the systen is $2d = N\delta$ where $N$ is the number of layers with negative 
dielectric constant. } 
\label{gain0_fig} 
\end{figure} 
 
Now let us consider the layered system shown in Fig. ({\ref{gain0_fig}), but 
with the gain included in the positive medium so as to exactly counter the 
effects of absorption in the negative medium. This can, for example, be 
accomplished by using a semiconductor laser material such as GaN or AlGaAs 
for the positive medium and silver for the negative medium. Using 
blue/ultra-violet (UV) light to pump the AlGaAs, (silver being transparent 
to the UV light), one can make the AlGaAs now optically amplifying in the 
red region of the spectrum, where one can satisfy the perfect lens condition 
for the real parts of the dielectric constant. By adjusting the pump laser 
intensity, the imaginary part of the positive gain medium can be tuned. The 
imaging can now be carried out in the red. Of course, it would be possible 
to use other materials and correspondingly different wavelengths of light. 
Alternatively one could also use other high gain processes such as Raman gain 
for this purpose.  } 
 
\begin{figure}[tbp] 
\includegraphics[width=8cm]{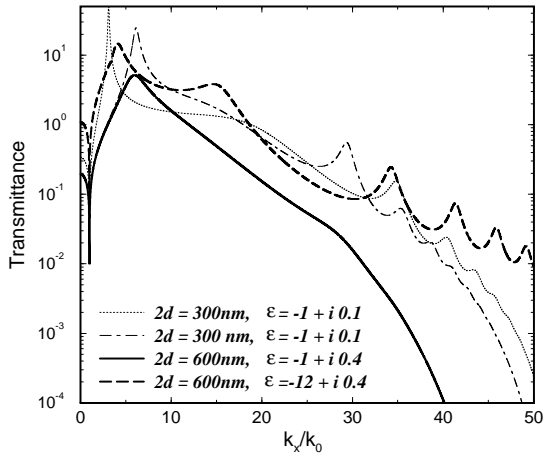} 
\caption{The transmission coefficient for a layered stack of positive and 
negative dielectric layers when the positive media are optically amplifying. 
The thicknesses of the positive and the negative dielectric layers are $%
\delta /2=$ 10nm. } 
\label{gain1_fig} 
\end{figure} 
We show the transmission coefficient, calculated using the transfer matrix 
method\cite{bornwolf}, for this layered medium with gain in Fig.~(\ref 
{gain1_fig}). Our numerical calculations are exact and are not  
carried out in the near-field approximation.  
The transmission is large and comparatively flat (compared to the lossy passive
case), almost up to 
the point ($k_x$) where the transmission resonances for the corresponding 
lossless system occurs \cite{anantha02a,anantha02b} and the transmission 
begins to decay exponentially. It is almost as if the effects of absorption 
have been cancelled out. Further, it is almost independent of the 
total length of the stack. However, it does not result in an exact 
cancellation and this can be seen from the fact that the transmission 
resonances of a completely lossless (and gainless) system are not restored. 
One can see some remnants of these resonances for the case when $\varepsilon 
_{-}=-1+i0.1$, and even lesser for the more absorptive (realistic for 
silver) case of $\varepsilon _{-}=-1+i0.4$. But it is clear that the image 
resolution will be improved vastly and to an almost similar extent in both 
cases, regardless of the levels of absorption/gain, provided the gain is 
large enough to compensate for the absorption. Thus with amplification 
included, the deleterious aspect of absorption, that it limits resolution, 
is cancelled out while the desirable aspect, that it softens the 
transmission resonances by preventing divergences, is retained. 
Only the transmission resonance close to $k_0$ is relatively undamped.
In Fig.~(\ref%
{gain1_fig}), the transmission across a multilayer stack of 600 nm thickness 
is shown (the individual layers are of 10nm thickness). We see that the 
transmission reduces only to 0.1 at $k_x/k_0\sim 20$ for $\varepsilon 
_{-}=-1+i0.4$ and an image with a resolution of about $\lambda/20$ can, 
in principle,  be 
transferred across the distance of the order of $\lambda$ (the wavelength of 
light). Note that the resolution can be marginally larger when the 
small transmission tail beyond is considered. 
The case when a high index dielectric such 
as AlGaAs is used and the wavelength is tuned to match the perfect lens 
conditions $\varepsilon _{\pm }=\pm 12\mp i0.4$, ($\lambda \simeq $ 578 nm 
for silver), is also shown in Fig.~(\ref{gain1_fig}). Although it appears 
from the graph that the degree of sub-wavelength resolution is larger, using 
a large dielectric constant would give slightly lower absolute resolution  
as the wavelength is larger. We should also note that making the 
layers thinner would improve the resolution even further, and we have been 
quite conservative in our choice of the layer thickness. 
 
\begin{figure}[tbp] 
\includegraphics[width=8cm]{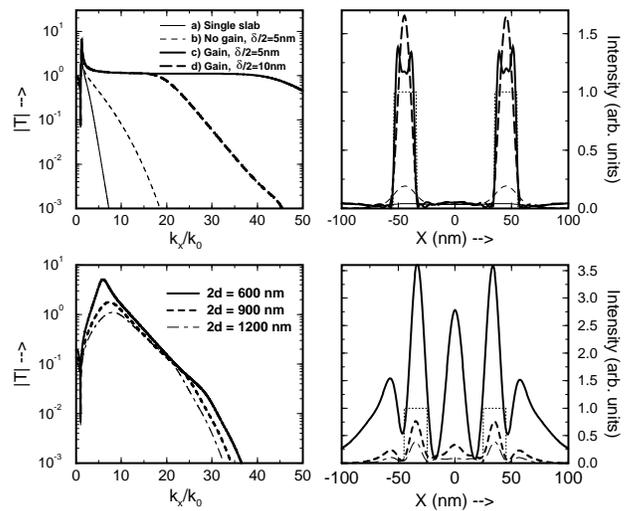} 
\caption{Top panels: The transmission function (left) and the electromagnetic
field intensity at the image plane (right) obtained (a) with a single slab 
of silver of 40 nm thickness, (b) When the slab is split into 8 thin layers 
of $\delta/2 = $ 5nm thicknesses (c) Layered silver-dielectric 
stack with optical gain and $\delta/2 = $ 5nm and (d) Layered silver-dielectric
stack with optical gain and $\delta/2 = $ 10 nm. 
$\varepsilon_{\pm} = \pm 1 \mp i 0.4$ in (c) and (d). 
Bottom panels:  
For a layered stack of large total thickness with optical 
gain($\varepsilon_{\pm} = \pm 1 \mp 0.4$) and $\delta/2 =$ 10nm for $2d=$ 
600nm, 900nm and 1200nm. The dotted lines show the position of the slit in 
the object plane} 
\label{gain2_fig} 
\end{figure} 
We show in Fig. (\ref{gain2_fig}), the transmission as a function of the
parallel wave vector and the images obtained by such layered media 
with optical gain. The object consists of  two slits of 20 nm width which 
are separated by 80 nm peak-to-peak distance. 
In the top panel we show the images for 
the case when the distance between the object plane to the image plane is 
$2d=$ 80 nm. For 
comparision, we also show the case of the original single slab of silver as the 
lens (solid line and $\delta/2=$ 40 nm) and a layered but gainless system. 
The two peaks in the image for the single slab can hardly be resolved, 
while they are clearly resolved in the case of the layered system with no gain.
The improvement in the image resolution for the layered system with 
gain over the corresponding gainless systems is obvious with the sharp
edges of the slits becoming visible. 
In the bottom panel we show the images formed by layered media with gain but 
with very large total thicknesses of the order of few wavelengths ($2d =$ 
600nm, 900nm and 1200nm). For a lossy passive system, in comparision, almost 
nothing would be visible at such large distances from the source. 
In the case of the 
layered system with gain, although the slits in the image are well resolved, 
there are extra background structures that show up in the image  
corresponding to the larger transmission at smaller $k_x$. 
There is 
overcompensation in the lens for (subwavelength) wave vectors in the range ($%
k_{0} < k_{x}<10 k_{0}$) due to the band of transmission resonances that are
relatively undamped near $k_0$, 
and hence, the transmission function is not 
constant with the wave-vector.  This overcompensated amplification also results in the large image intensity. We note that the effect of these resonances 
reduces for much larger thicknesses and the transmission function is actually 
more constant for thicker stacks.  
Although waves with very large wave vectors also get effectively amplified
even for large stack thicknesses,
the transmission resonances do not allow a clean image to be produced. 
We note, however, that since the high spatial frequency components are transferred
across, knowledge of the transmission function of the lens 
would enable one to recover a clean image from the observed image.
Alternatively, structuring the silver slabs in the transverse direction
will create plasmonic bandstructures and one can engineer these band-structures
to inhibit the transmission resonances close to $k_0$.
The transmission functions suggest that an image 
resolution of almost $\lambda/25$ could easily be achieved with these systems. 

As a note of caution, we note that in the presence of intense field 
enhancements that are expected in this system the gain is likely to get 
saturated. The largest field enhancements will be for the largest transverse 
wave-vectors. If we make the layers very thin, the local field enhancements 
will not be as intense and the gain might not get completely bleached. In 
general, however, we do expect that the effective gain will be somewhat 
reduced and the corresponding enhancements of the image resolution will be 
smaller. Another point of concern is that the system with gain could become 
unstable and undergo self-sustaining oscillations in the transverse 
direction due to spontaneous symmetry breaking. The layers would now just 
act as wave guides. This would bleach the system of all gain and the whole 
effect could be lost. It is clear that such effects will be minimized for a 
system with very thin layers. We are analyzing these effects which are 
beyond the scope of this paper and those results will be presented elsewhere. 
 
In conclusion, we have shown that a multilayer stack of thin alternating 
layers of silver and a positive amplifying dielectric medium with optical 
gain/amplification can transport evanescent waves with very little 
attenuation even over large stack thicknesses. This novel \textit{optical 
fibre bundle} can thus act on the evanescent near-field of a radiating 
source and images with high sub-wavelength resolution can be transferred 
across. The potential applications are immense and range from nanoscale 
lithography to near-field optical imaging and data storage.  
 
\begin{acknowledgments}  
SAR would like to thank Prof.  Geoff New for discussions and acknowledge 
the support from DoD/ONR MURI grant N00014-01-1-0803. 
\end{acknowledgments}  

\end{document}